# Competitive and Weighted Evolving Simplicial Complexes

Zhaohua Guo[1], Rui Miao[1**], Jin-Li Guo[2,3**], Yuan Yuan[3], Jeffrey Yi-Lin Forrest[4]

[1]School of Ocean and Civil Engineering, Shanghai Jiao Tong University, 200240, China;
[2]Department of Data Science and Engineering, Xi'an Innovation College of Yan'an University, 710100，China;
[3]Business School, University of Shanghai for Science and Technology, Shanghai, 200093, China;
[4]Department of Accounting Economics Finance, Slippery Rock University, Slippery Rock, PA 16057, USA

**Abstract**: A simplex-based network is referred to as a higher-order network, in which describe that the interactions can include more than two nodes. Many multicomponent interactions can be grasped through simplicial complexes, which have recently found applications in social, technological, and biological contexts. The paper first proposes a competitive evolving model of higher-order networks. We introduce the difference equation analysis approach in the high-order network to make the analysis network more rigorous. It avoids the assumption that the degrees of nodes are continuous in the traditional analysis network. We obtain an analytical expression for the distribution of higher-order degrees by employing the theory of Poisson processes. The established results indicate that in a $d$-order network the scale-free behavior for the ($d$-1)-dim simplex with respect to the $d$-order degree is controlled by the competitiveness factor. As the competitiveness increases, the $d$-order degree of the ($d$-1)-dim simplex is bent under the logarithmic coordinates. While the $e(< d-1)$-dim simplex with respect to the $d$-order degree exhibits scale-free behavior. Second, by considering the weight changes of the neighboring simplices, as triggered by the selected simplex, a new weighted evolving model in higher-order networks is proposed. The results of the competitive evolving model of higher-order networks are used to analyze the weighted evolving model so that obtained are the analytical expressions of the higher-order degree distribution and higher-order strength density function of weighted higher-order networks. The outcomes of the simulation experiments are consistent with the theoretical analysis.

---

[**] Corresponding author. Rui Miao. Email: miaorui@sjtu.edu.cn
[**] Corresponding author. Jin-Li Guo. Email: phd5816@163.com

## 1. Introduction

Network is everywhere. Before the arrival of this century, people often employed ER model to describe real-life networks. Since the time when Watts and Strogatz [1] developed the small-world model at the end of the 20$^{th}$ century and Barabási and Albert [2] proposed the scale-free network model, an upsurge in studies of complex network has been set off. In the past twenty plus years, the research on complex networks has been very active, involving scholars from many different fields, such as physics, computer science, graph theory, transportation, economics, and others. The relevant researches range from Internet networks to WWW networks, from power networks to transportation networks, from scholarly collaboration networks to various economic, political and social relationship networks. Through over twenty some years of efforts, scholars have constructed a large number of network models and introduced many methods to analyze these models. The theoretical research on complex networks has gradually transitioned from extensive research at the end of the last century to rigorous research, and the relevant theory has gradually matured.

The complex network theory has become a powerful tool for studying complex systems. The past two decades have seen significant successes in our understanding of networked systems. However, the complex network is restricted to pairwise interactions. With the development of society and the advancement of science and technology, the complexity of real-life systems under consideration continuously increases so that complex networks can no longer fully describe such real-life networks. For example, it is extremely difficult to describe by using complex networks such sophisticated multi-interactive systems as (i) that consisting of WeChat groups as edges and group members as nodes, (ii) that consisting of family members as nodes and each family as an edge, (iii) that consisting of airlines as nodes, and transportation airports as edges.

With the development of big data technology, higher-order networks that are able to depict multiple interactions have emerged in recent years. The research of higher-order networks has gradually attracted scholars' attention [3-14]. By higher-order networks, we mean hypergraphs and simplicial complexes [8]. Boccaletti et al. [3] discussed the relationship between hypergraphs and simplicial complexes. The hypergraph-based definition of higher-order networks is straightforward [9], the concept such defined cannot be investigated by employing such tools as algebraic topology. In the contrary, although the definition of higher-order networks based on simplices is relatively abstract, the tools of algebraic topology can be used to describe the higher-order structure of the network, we use this definition to describe the higher-order structure of networks by using algebraic topology as a tool. Empirical research on higher-order networks mainly focuses on confirming the network characteristics of real systems. For example, Hu Feng and others constructed a higher-order network using proteins as nodes and complexes as hyperedges, analyzed the structural characteristics of the network, and identified key proteins [10]; In reference [11], authors constructed a bus hypernetwork by regarding bus stations as hyperedges and bus routes as nodes, and analyzed its network characteristics. By analyzing the robustness of the network, they concluded that the stability of random attacks is stronger than that of deliberate attacks [11]. Guo and Suo take into account the brand effect and competitiveness in hypernetworks, and propose a higher-order network evolution model based on hypergraphs [9].

In terms of evolution models of higher-order networks based on simplices, an important model is the so-called "network geometry with flavor" (NGF) model [15]. This model not only generalizes the BA model of complex networks to simplicial complexes, but also generates simplicial complexes of any dimension $d$ with very distinct statistical and combinatorial properties. The displayed hyperbolic geometry can be readily explored. The original NGF model adds one maximal simplices of fixed dimension at each time step of the growth process. Kovalenko et al. [7] introduced the generalized model that adds more maximal simplices of fixed dimension at each time step. Fountoulakis et al.[13] introduced even the option to remove an existing simplex when a new one is added. Although these evolution models can describe higher-order interactions, they are unweighted simplicial complexes and cannot reflect the competitive nature of higher-order interactions. For example, the collaborative network of scientists without any weight does not reflect the number of joint papers. Courtney and Bianconi established a growing weighted simplicial complex model [5]. At each time step, this weighted network adds one new node, selects several ($d$-1)-dim simplices to form several $d$-dim simplices, and at the same time, selects several $d$-dim simplexes in the network, and increases the weight $w_0$ of each selected $d$-dim simplex. Although this model well describes the growing weighted simplicial complex, it cannot reflect the change in the weights of neighboring edges, as stimulated by the ($d$-1)-dim simplices selected by new node and cannot also degenerate into the BBV weighted network in complex networks [16].

The main purpose of this paper is to establish a competitive simplicial complex model and a simplicial complex model that adds new simplex nodes to stimulate changes in the weights of neighboring edges.

This paper is organized as follows. The next section introduces the concept of simplex-based higher-order networks and the higher-order strength. The third section establishes a competitive evolving model of higher-order networks, which is then analyzed by using the difference equation approach and Poisson process theory. Doing so leads to an analytical expression of the distribution of higher-order degrees. And the model is simulated on a computer. The fourth section develops a new weighted evolving simplicial complex model. The properties of competitive simplicial complexes are used to obtain the higher-order degree distribution and higher-order strength density function of the weighted simplicial complexes. Then simulation experiments are conducted on the model. Our analysis and simulation show that this model embodies the emergence of scaling laws of higher-order degrees of the low-dimensional simplex in higher-order networks. The fifth section summarizes the entire work presented in this paper.

## 2. The Concept of Higher-order Networks

The definition of higher-order network is based on simplices: Assume that $V = \{v_1, v_2, \cdots, v_n\}$ is a finite set. If as a subset α of $V = \{v_1, v_2, \cdots, v_n\}$, $\alpha = [v_{\alpha,0}, v_{\alpha,1} \cdots, v_{\alpha,d-1}, v_{\alpha,d}] \neq \Phi$, then α is known as a $d$-dim simplex [15]. A subset $\alpha'$ of α (that is $\alpha' \subset \alpha$) is referred to as a face of simplex α. $K$ is known as a simplicial complex based on $V = \{v_1, v_2, \cdots, v_n\}$, if its elements are subsets of $V$ satisfying

1. If $\alpha$ is a simplex in $K$, then each arbitrary face of $\alpha$ belongs to $K$.
2. If $s_1$ and $s_2$ are two arbitrary simplexes in $K$, then $s_1 \cap s_2$ is empty or one of $s_1$ and $s_2$ is a common face.

The simplicial complex $K$ is called a higher-order network based on simplices. The maximum dimension number of simplices in $K$ is known as the dimension of the simplicial complex, denoted $dimK$. $K$ is also known as an $dimK$-order network. The $d$-dim higher-order degree $h_{d,m}^{\alpha}$ of an $m$-dim simplex $\alpha$ (abbreviated as the $d$-order degree) stands for the number of $d$-dim simplices including $\alpha$. The $d$-dim higher-order degree $h_{d,d-1}^{\alpha}$ of $d$-1-dim simplex $\alpha$ is known as the sub-higher-order degree.

If $dimK \equiv 1$, the higher-order network degenerates into a complex network. Therefore, the concept of simplex-based higher-order networks is a generalization of that of complex networks.

A weighted higher-order network is an ordered triplet $(V, K, W)$, where $V$ is the set of nodes, $K$ a $d$-dim simplicial complex based on $V$, and $W$ the set of weights of the $d$-dim simplices in $K$.

Let

$$S_d = \{\alpha \in K | \alpha \text{ is a } d \text{ dim simplex}\}$$

$w_\beta \in W$ stands for the weight of $d$-dim simplex $\beta$. Define the $d$-dim higher-order strength of $n$-dim simplex $\alpha$, simply known as $d$-order strength, as follow:

$$s_{d,n}^\alpha = \sum_{\beta \in S_d | \beta \supseteq \alpha} w_\beta$$

The $d$-dim higher-order strength of $(d-1)$-dim simplex $\alpha$ is simply known the sub-higher-order strength.

## 3. Competitive Evolving Model of Simplicial Complexes

In a realistic network, it is often the case that nodes and groups of nodes possess certain competitiveness. For example, in the higher-order network of scientists' collaborations individual scholars are nodes, the authors of a research paper form a simplex. When new authors join the network, the fame of a scientist possesses additional competitiveness. Therefore, a famous scientist has a greater probability of being selected. In this section, we develop a model to describe such an evolutionary higher-order network with competitive simplexes. A simplicial complex evolution model is referred to as having competitive attraction, if it satisfies the following two conditions:

(1) *Growth*: It starts at $t = 0$ from an initial finite simplicial complex that comprises $m_0$ $d$-dimensional simplices, and every $(d-1)$-dim simplex has $d$-order degree at least one. The arriving process of new nodes is a Poisson process of constant rate $\lambda$. At time $t$ when a new node arrives the network, $m \leq (d+1)m_0$ new $d$-dim simplices formed by this new node and $m$ existing $(d-1)$-dim simplices in the network are added to the network;

(2) *Preferential attachment*: The probability $\Theta$ that the new $d$-dim simplex is glued to a $(d-1)$-dim simplex $\alpha$ depends on $d$-order degree $h_{d,d-1}^\alpha$ such that

$$\Theta(h_{d,d-1}^\alpha) = \frac{h_{d,d-1}^\alpha + a}{\sum_{\beta \in S_{d-1}(t)} (h_{d,d-1}^\beta + a)}. \tag{1}$$

where $a (> -1)$ stands for the competitiveness factor of simplex $\alpha$, $S_{d-1}(t)$ the set of all $(d-1)$-dim simplices in the network at time $t$.

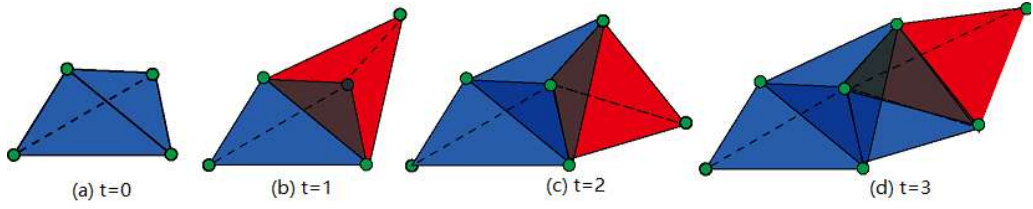

(a) t=0    (b) t=1    (c) t=2    (d) t=3

**Fig. 1**. Schematic illustration of the first three steps of the evolution for the model.

As can be seen from Fig. 1, at each time step we add a tetrahedron (3-dimensional simplex), resulting in an increase of 3 triangles (2-dimensional simplex) and 3 line segments (1-dimensional simplex).

Let $N(t)$ = the number of nodes in the network at time $t$ and $\mu(t) = E[N(t)]$.

Because the arriving process $N(t)$ of nodes is a Poisson process with constant rate $\lambda$, from the theory of Poisson processes, it follows have $E[N(t)] = \lambda t$. Let $t_\alpha$ denote the time when the last node of simplex $\alpha$ enters the network, $h^\alpha_{d,d-1}(N(t))$ the $d$-order degree of the ($d$-1)-dimensional simplex $\alpha$ at time $t$. Since the rate of change of $h^\alpha_{d,d-1}(N(t))$ is proportional to the probability $\Theta(h^\alpha_{d,d-1}(N(t)))$, $h^\alpha_{d,d-1}(N(t))$ satisfies the following equation.

$$h^\alpha_{d,d-1}(N(t)) - h^\alpha_{d,d-1}(N(t)-1) = m \frac{h^\alpha_{d,d-1}(N(t)-1)+a}{\sum_\beta (h^\beta_{d,d-1}(N(t)-1)+a)}. \tag{2}$$

The factor $m$ on the right-hand side reflects the possibility that simplex $\alpha$ could be respectively selected from the first until the $m$th time.

The second sum in the denominators of equation (2) is the total number of ($d$-1) faces in the simplicial complex at time $t$. At each timestep, we add $m$ new $d$-dim simplices, each of which contributes $d$ new ($d$-1) faces (ignoring possible repetitions among the new simplices). Hence, for $t \gg 1$, the number of ($d$-1) faces can be approximated as follows:

$$\sum_{\beta \in S_{d-1}(t)} 1 = c_{m,1}(d-1)dN(t) .$$

where $c_{m,1}(e) = \begin{cases} 1, & \text{if } e=0 \\ m, & \text{if } e>0 \end{cases}$

For $t \gg 1$, the first sum in the denominator of equation (2) is

$$\sum_{\beta \in S_{d-1}(t)} h^\beta_{d,d-1} = m(d+1)N(t) .$$

Therefore, we have

$$\sum_{\beta \in S_{d-1}(t)} (h^\beta_{d,d-1} + a) = m(d+1)N(t) + ac_{m,1}(d-1)dN(t) . \tag{3}$$

Substituting equation (3) into equation (2) produces

arXiv:2402.06451v2$$h_{d,d-1}^\alpha(N(t)) - h_{d,d-1}^\alpha(N(t)-1) = \frac{h_{d,d-1}^\alpha(N(t)-1)+a}{\left(\left(\frac{a}{m}c_{m,1}(d-1)+1\right)d+1\right)N(t)}. \quad (4)$$

For $e < d-1$, the probability for the $d$-order degree of an $e$-dim simplex $\delta$ to increase is

$$\Theta_e(\delta) = \sum_{\alpha \in S_{d-1}|\alpha \supseteq \delta} \Theta(h_{d,d-1}^\alpha) = \frac{\sum_{\alpha \in S_{d-1}|\alpha \supseteq \delta}(h_{d,d-1}^\alpha(N(t)-1)+a)}{\sum_{\beta \in S_{d-1}(t)}(h_{d,d-1}^\beta(N(t)-1)+a)}. \quad (5)$$

where

$$\sum_{\alpha \in S_{d-1}|\alpha \supseteq \delta}(h_{d,d-1}^\alpha + a) = \sum_{\alpha \in S_{d-1}|\alpha \supseteq \delta} h_{d,d-1}^\alpha + a \sum_{\alpha \in S_{d-1}|\alpha \supseteq \delta} 1 \quad (6)$$

From reference [5], it follows that for $e_1 > e$,

$$h_{d,e}^\beta = \frac{1}{\binom{d-e}{e_1-e}} \sum_{\alpha \in S_{e_1}|\alpha \supseteq \beta} h_{d,e_1}^\alpha \quad (7)$$

From equation (7), it follows that

$$\sum_{\alpha \in S_{d-1}|\alpha \supseteq \delta} h_{d,d-1}^\alpha = (d-e)h_{d,e}^\delta \quad (8)$$

Since $h_{d,e}^\delta(t)$ stands for the $d$-order degree of $e$-dim simplex $\delta$ at time $t$, from reference [17] it follows that

$$\sum_{\alpha \in S_{d-1}|\alpha \supseteq \delta} 1 = h_{d-1,e}^\delta = \begin{cases} m + (d-1)h_{d,e}^\delta, & \text{if } e = 0 \\ 1 + (d-e-1)h_{d,e}^\delta, & \text{if } e > 0 \end{cases} = c_{1,m}(e) + (d-e-1)h_{d,e}^\delta, \quad (9)$$

where $c_{1,m}(e) = \begin{cases} m, & \text{if } e = 0 \\ 1, & \text{if } e > 0 \end{cases}$.

Substituting equations (8) and (9) into equation (6) leads to

$$\sum_{\alpha \in S_{d-1}|\alpha \supseteq \delta} (h_{d,d-1}^\alpha + a)$$
$$= (d-e)h_{d,e}^\delta + a \begin{cases} m + (d-1)h_{d,e}^\delta, & \text{if } e = 0 \\ 1 + (d-e-1)h_{d,e}^\delta, & \text{if } e > 0 \end{cases}$$
$$= ac_{1,m}(e) + \big((a+1)(d-e) - a\big)h_{d,e}^\delta. \quad (10)$$

Since the rate of change of $h_{d,e}^\delta(t)$ is proportional to the probability $\Theta_e(\delta)$, we know that $h_{d,e}^\delta(t)$ satisfies the following equation

$$h_{d,e}^{\delta}(N(t)) - h_{d,e}^{\delta}(N(t) - 1) = m \frac{\sum_{\alpha \in S_{d-1} | \alpha \supseteq \delta} \left( h_{d,d-1}^{\alpha}(N(t) - 1) + a \right)}{\sum_{\beta \in S_{d-1}(t)} \left( h_{d,d-1}^{\beta}(N(t) - 1) + a \right)}. \quad (11)$$

By substituting equations (3) and (10) into equation (11), we have

$$h_{d,e}^{\delta}(N(t)) - h_{d,e}^{\delta}(N(t) - 1) = \frac{ac_{1,m}(e) + [(a+1)(d-e) - a]h_{d,e}^{\delta}(N(t) - 1)}{\left( \left( \frac{a}{m} c_{m,1}(d-1) + 1 \right) d + 1 \right) N(t)}. \quad (12)$$

Case 1  $e \neq d - \frac{a}{a+1}$

$$ac_{1,m}(e) + [(a+1)(d-e) - a]h_{d,e}^{\delta}(N(t))$$

$$= \left( 1 + \frac{(a+1)(d-e) - a}{\left( \left( \frac{a}{m} c_{m,1}(d-1) + 1 \right) d + 1 \right) N(t)} \right) \{ ac_{1,m}(e) + [(a+1)(d-e) - a]h_{d,e}^{\delta}(N(t) - 1) \}. \quad (13)$$

Let $\gamma = \frac{\left( \frac{a}{m} c_{m,1}(d-1) + 1 \right) d + 1}{(a+1)(d-e) - a}$, then

$$\ln \frac{ac_{1,m}(e) + [(a+1)(d-e) - a]h_{d,e}^{\delta}(N(t))}{ac_{1,m}(e) + [(a+1)(d-e) - a]h_{d,e}^{\delta}(N(t_{\delta}))} = \sum_{j=N(t_{\delta})+1}^{N(t)} \ln \left( 1 + \frac{1}{\gamma j} \right). \quad (14)$$

By ignoring the calculations of connecting to neighboring faces, therefore,

$$h_{d,e}^{\delta}(N(t_{\delta})) = c_{1,m}(e) = \begin{cases} m, & \text{if } e = 0 \\ 1, & \text{if } e > 0 \end{cases}, \quad (15)$$

$$\ln \frac{ac_{1,m}(e) + [(a+1)(d-e) - a]h_{d,e}^{\delta}(N(t))}{ac_{1,m}(e) + [(a+1)(d-e) - a]c_{1,m}(e)} \approx \frac{1}{\gamma} \left( \sum_{j=1}^{N(t)} \frac{1}{j} - \sum_{j=1}^{N(t_{\delta})} \frac{1}{j} \right). \quad (16)$$

Since $E[N(t)] = \lambda t$, $E[N(t_{\delta})] = \lambda t_{\delta}$ and

$$\sum_{j=1}^{N(t)} \frac{1}{j} - \sum_{j=1}^{N(t_{\delta})} \frac{1}{j} = C + \ln N(t) + \varepsilon_{N(t)} - [C + \ln N(t_{\delta}) + \varepsilon_{N(t_{\delta})}], \quad (17)$$

where, $C$ is Euler's constant, and $\lim_{n \to \infty} \varepsilon_n = 0$, we have

$$\frac{ac_{1,m}(e) + [(a+1)(d-e) - a]h_{d,e}^{\delta}(N(t))}{ac_{1,m}(e) + [(a+1)(d-e) - a]c_{1,m}(e)} \approx \left(\frac{t}{t_\delta}\right)^{\frac{1}{\gamma}}.$$

Therefor,

$$h_{d,e}^{\delta}(N(t)) \approx \frac{\left(ac_{1,m}(e) + [(a+1)(d-e) - a]c_{1,m}(e)\right)}{[(a+1)(d-e) - a]}\left(\frac{t}{t_\delta}\right)^{\frac{1}{\gamma}} - \frac{ac_{1,m}(e)}{[(a+1)(d-e) - a]}. \quad (18)$$

From Equation (18), we have

$$P\{h_{d,e}^{\delta}(N(t)) \geq k\} = P\left\{t_\delta \leq \left(\frac{(a+1)(d-e)c_{1,m}(e)}{[(a+1)(d-e) - a]k + ac_{1,m}(e)}\right)^\gamma t\right\}. \quad (19)$$

Without loss of generality, let $\delta = n(n = 0,1,2, \dots,)$ denote the $e$-dimensional simplex of the n*th* batch entering the network, then

$$P\{h_{d,e}^{n}(N(t)) \geq k\} = P\left\{t_n \leq \left(\frac{(a+1)(d-e)c_{1,m}(e)}{[(a+1)(d-e) - a]k + ac_{1,m}(e)}\right)^\gamma t\right\}. \quad (20)$$

By Poisson process theory [18], $t_0 = 0, t_1 - t_0, t_2 - t_1, \dots, t_n - t_{n-1}$ are independent and identically distributed exponential random variables having mean $1/\lambda$. the arrival time $t_n = t_n - t_{n-1} + t_{n-1} - t_{n-2} + \cdots + t_2 - t_1 + t_1 - t_0$ of node $n$ follows a gamma distribution with parameters $n$ and $\lambda$.

$$P(t_n \leq x) = e^{-\lambda} \sum_{l=n}^{\infty} \frac{1}{l!}(\lambda x)^l = 1 - e^{-\lambda} \sum_{l=0}^{n-1} \frac{1}{l!}(\lambda x)^l.$$

Let $F_n(x)$ denote the *n*th convolution of the exponential distribution with the parameter $\lambda$, then

$$P\left\{t_n \leq \left(\frac{(a+1)(d-e)c_{1,m}(e)}{[(a+1)(d-e)-a]k+ac_{1,m}(e)}\right)^\gamma t\right\} = F_n\left(\left(\frac{(a+1)(d-e)c_{1,m}(e)}{[(a+1)(d-e)-a]k+ac_{1,m}(e)}\right)^\gamma t\right). \quad (21)$$

Since

$$\sum_{n=1}^{+\infty} F_n\left(\left(\frac{(a+1)(d-e)c_{1,m}(e)}{[(a+1)(d-e)-a]k+ac_{1,m}(e)}\right)^\gamma t\right) = E\left[N\left(\left(\frac{(a+1)(d-e)c_{1,m}(e)}{[(a+1)(d-e)-a]k+ac_{1,m}(e)}\right)^\gamma t\right)\right]. \quad (22)$$

Therefore, we know that the *d*-order cumulative distribution of the *e*-dimensional simplex is

$$P_{d,e}^{cum}(k) \approx \lim_{t \to \infty} \frac{1}{E[N(t)]} \sum_{n=1}^{\infty} F_n\left(\left(\frac{(a+1)(d-e)c_{1,m}(e)}{[(a+1)(d-e)-a]k+ac_{1,m}(e)}\right)^\gamma t\right)$$

$$= \left(\frac{(a+1)(d-e)c_{1,m}(e)}{[(a+1)(d-e)-a]k+ac_{1,m}(e)}\right)^\gamma. \quad (23)$$

Where

$$\gamma = \frac{\left(\frac{a}{m}c_{m,1}(d-1)+1\right)d+1}{(a+1)(d-e)-a}. \quad (24)$$

From Equation (23), the distribution of the *d*-degree of *e*(<*d*)-dimensional simplices $\delta$ is given by

$$P_{d,e}(k) = \gamma \frac{(a+1)(d-e)-a}{(a+1)(d-e)c_{1,m}(e)}\left(\frac{(a+1)(d-e)c_{1,m}(e)}{[(a+1)(d-e)-a]k+ac_{1,m}(e)}\right)^{\gamma+1}. \quad (25)$$

If $e = d - \frac{a}{a+1}$

$$h_{d,e}^{\delta}(N(t)) = \frac{ac_{1,m}(e)}{\left(\left(\frac{a}{m}c_{m,1}(d-1)+1\right)d+1\right)N(t)} + h_{d,e}^{\delta}(N(t)-1). \quad (26)$$

Let $b = \left(\frac{a}{m}c_{m,1}(d-1)+1\right)d+1$, then

$$h_{d,e}^{\delta}(N(t)) = \frac{ac_{1,m}(e)}{b} \sum_{j=N(t_\delta)+1}^{N(t)} \frac{1}{j} + h_{d,e}^{\delta}(N(t_\delta)). \quad (27)$$

From Equation (17), we have

$$h_{d,e}^{\delta}(N(t)) = \frac{ac_{1,m}(e)}{b}\left(\sum_{j=1}^{N(t)}\frac{1}{j} - \sum_{j=1}^{N(t_{\delta})}\frac{1}{j}\right) + c_{1,m}(e). \tag{28}$$

From Equation (17) and Equation (28), we have

$$h_{d,e}^{\delta}(N(t)) = \frac{ac_{1,m}(e)}{b}\ln\frac{N(t)}{N(t_{\delta})} + c_{1,m}(e). \tag{29}$$

Therefore, we have

$$h_{d,e}^{\delta}(N(t)) = \frac{ac_{1,m}(e)}{\left(\frac{a}{m}c_{m,1}(d-1)+1\right)d+1}\ln\frac{t}{t_{\delta}} + c_{1,m}(e). \tag{30}$$

Without loss of generality, let $\delta = n(n = 0,1,2,\dots,)$ denote the $e$-dimensional simplex of the n*th* batch entering the network, from Equation (30), we have

$$P\{h_{d,e}^{n}(N(t)) \geq k\} = P\left\{t_n \leq te^{-\frac{\left(\frac{a}{m}c_{m,1}(d-1)+1\right)d+1}{ac_{1,m}(e)}(k-c_{1,m}(e))}\right\}. \tag{31}$$

Similarly to the case $e \neq d - \frac{a}{a+1}$, we obtain that

$$P_{d,e}^{cum}(k) = e^{-\frac{\left(\frac{a}{m}c_{m,1}(d-1)+1\right)d+1}{ac_{1,m}(e)}(k-c_{1,m}(e))}. \tag{32}$$

From Equation (32), the distribution of the $d$-degree of $e(<d)$-dimensional simplices $\delta$ is given by

$$P_{d,e}(k) = \frac{\left(\frac{a}{m}c_{m,1}(d-1)+1\right)d+1}{ac_{1,m}(e)}e^{-\frac{\left(\frac{a}{m}c_{m,1}(d-1)+1\right)d+1}{ac_{1,m}(e)}(k-c_{1,m}(e))}. \tag{33}$$

Equation (33) indicates that the simplicial complex network under such conditions does not have any scale-free characteristic.

It can be shown that equation (25) is a straightened distribution of the bending power law. When $d-e-\frac{a}{a+1} \to 0$, that is, $(a+1)(d-e)-a \to 0$, equation (25) approaches equation (33).

When $e=d-1$, if $a \to \infty$, then $\frac{a}{a+1} \to 1$, $d-e-\frac{a}{a+1} \to 0$ so that $P_{d,d-1}(k)$ approaches an exponential distribution, that is, equation (33). When $e<d-1$, if $a \to \infty$ and $d-e-\frac{a}{a+1} \neq 0$, then $P_{d,e}(k)$ is a power-law distribution, that is equation (25). Therefore, it can be seen that the scale-freeness of a ($d$-1)-dim simplex with respect to the $d$-order degree is controlled by the competitiveness factor. As the competitiveness factor increases, the $d$-order degrees of ($d$-1)-dim simplices are bent under logarithmic coordinates. At the same time, the scale-freeness of the $d$-order degrees of $e$ ($< d - 1$)-dim simplices is not influenced by the competitiveness factor.

Figs. 2 – 5 depict the simulated and theoretically predicted values of our model developed above for different competitiveness factor values. These figures show that simulated values agree well with the theoretical analysis of the simplicial complex network evolution model with competitive attractions. From Fig. 2 and Fig. 3, it can be seen that the distribution of the $d$-order degree of ($d$-1)-dim simplices is bent under the log–log coordinates, while the distribution of the $d$-order degree of $e(< d - 1)$-dim simplices is relatively straight. It can be seen that scale-free phenomena appear in low-dimensional simplexes in higher-order networks. The greater the competitive power, the more curved the distribution of the $d$-order degree of ($d$-1)-dim simplices is in a log-log coordinate system. The solid lines in Fig.4 and Fig.5 are the simulated values of the cumulative distribution of 2-order degrees of the model with $a = -0.5$. Therefore, the competitiveness factor in the model can be negative.

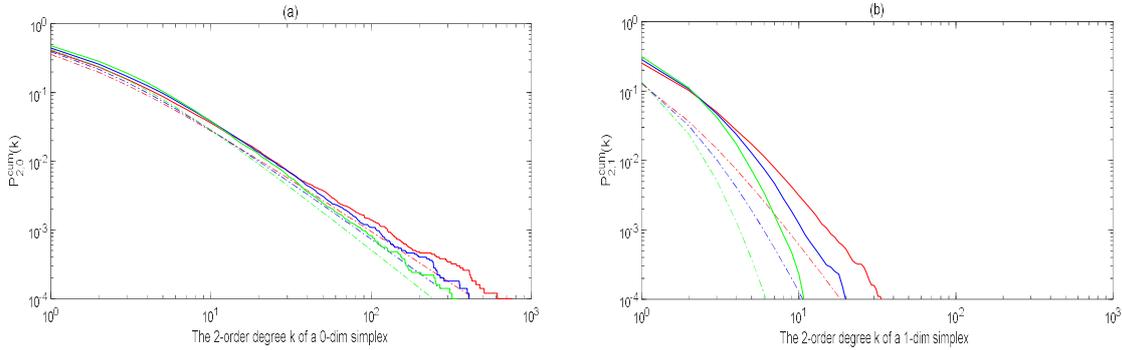

**Fig. 2 The cumulative distribution of the higher-order degree of the model with $m = 1$.** The cumulative distribution $P_{2,e}^{cum}(k)$ vs. the higher-order degree $k$ (number of triangles a link (node) $e$ is adjacent to) obtained by growing a network of size N = 50,000 with the preferential attachment Eq. (1). The lines in (a) stand for the cumulative distribution $P_{2,0}^{cum}(k)$ of the 2-order degree $k$ of the 0-dim simplex. The lines in (b) stand for the cumulative distribution $P_{2,1}^{cum}(k)$ of the 2-order degree $k$ of the 1-dim simplex. The solid lines in (a) and (b) are the simulated values of the cumulative distribution of 2-order degrees of the model. The dashed lines in (a) and (b) correspond to the analytical predictions given by Eq. (23). The red lines are the cumulative distribution of the 2-order degree of the model with the competitiveness factor $a = 0.1$. The blue lines are the cumulative distribution of the 2-order degree of the model with the competitiveness factor $a = 1$. The green lines are the cumulative distribution of the 2-order degree of the model with the competitiveness factor $a = 5$.

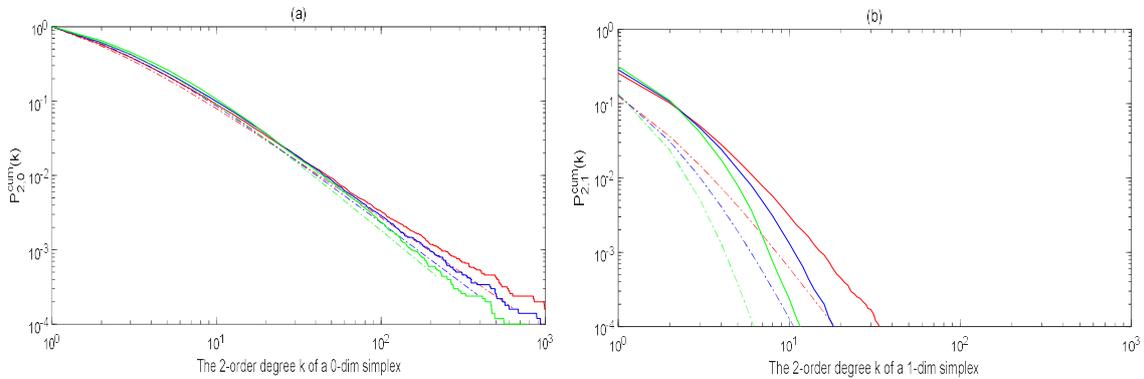

**Fig. 3 The cumulative distribution of the higher-order degree of the model with $m = 2$.** The rest of the caption is the same as that in Fig. 2.

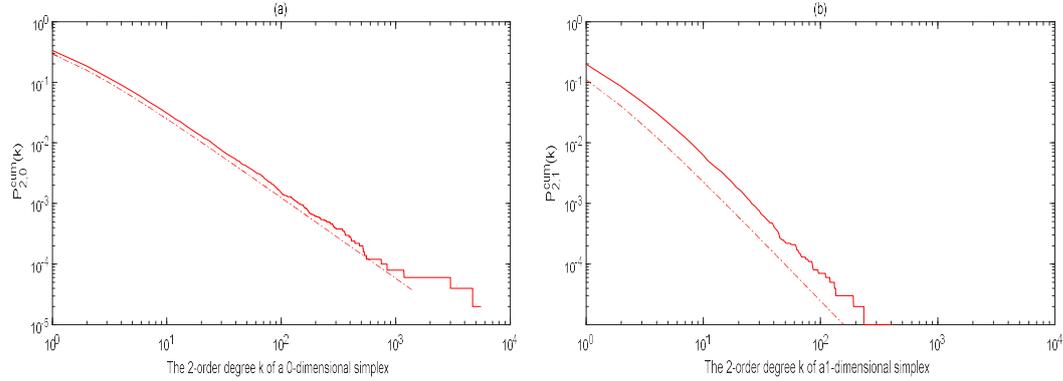

**Fig. 4. The cumulative distribution of the higher-order degree of the model with $m = 1$.** The cumulative distribution $P_{2,e}^{cum}(k)$ vs. the higher-order degree $k$ (number of triangles a link (node) $e$ is adjacent to) obtained by growing a network of size N = 50,000 with the preferential attachment Eq. (1). The lines in (a) stand for the cumulative distribution $P_{2,0}^{cum}(k)$ of the 2-order degree $k$ of the 0-dim simplex. The lines in (b) stand for the cumulative distribution $P_{2,1}^{cum}(k)$ of the 2-order degree $k$ of the 1-dim simplex. The solid lines in (a) and (b) are the simulated values of the cumulative distribution of 2-order degrees of the model. The dashed lines in (a) and (b) correspond to the analytical predictions given by Eq. (23). The red lines are the cumulative distribution of the 2-order degree of the model with the competitiveness factor $a = -0.5$.

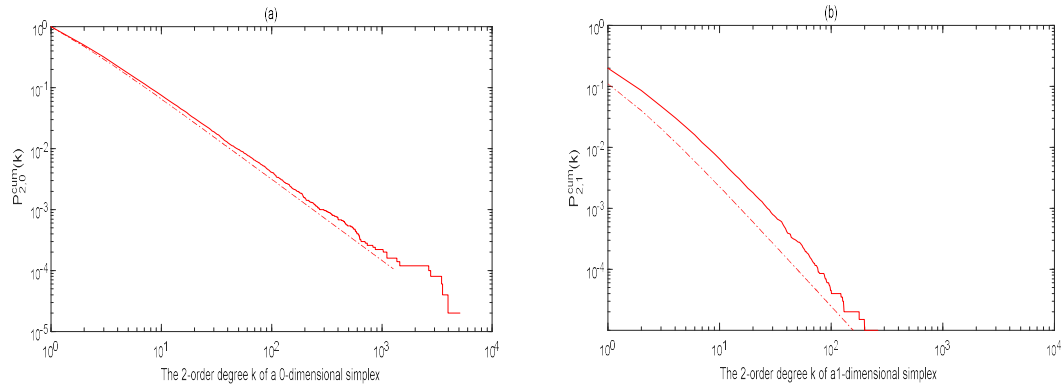

**Fig. 5. The cumulative distribution of the higher-order degree of the model with $m = 2$.** The rest of the caption is the same as that in Fig. 4.

## 4. Weighted Evolving Model of Simplicial Complexes

In real applications, often simplicial complexes are weighted. For instance, in scientific collaboration networks teams of collaborators can be weighted by the strength of the number of their collaboration. A weighted simplicial complex network evolution model is a higher-order network that satisfies the following three conditions:

(1) We start at time $t = 0$ from an initial finite simplicial complex that comprises $m_0$ $d$-dim simplices, and each $d$-dim simplex have weight $w_0$. The process for nodes to arrive the network is Poisson with constant rate $\lambda$. At time $t$, when a new node arrives the network, this new node and an existing ($d$-1)-dim simplex form a $d$-dim simplex with weight $w_0$, leading to $m(\leq (d+1)m_0)$ $d$-dim simplices without any repetitive non-zero dimensional simplices.

(2) When selecting an existing ($d$-1)-dim simplex $\alpha$ to connect with the new node, the probability $\Pi$ of selecting simplex $\alpha$ depends on the $d$-order strength $s_{d,d-d}^{\alpha}$ of $\alpha$, satisfying

$$\Pi(s_{d,d-d}^{\alpha}) = \frac{s_{d,d-d}^{\alpha}}{\sum_{\beta \in S_{d-1}(t)} s_{d,d-d}^{\beta}}. \tag{34}$$

where $S_{d-1}(t)$ stands for the set of all ($d$-1)-dim simplices in the network at time $t$, and $s_{d,d-d}^{\alpha} = \sum_{\beta \in S_d | \beta \supseteq \alpha} w_\beta$.

(3) The initial weight of each new $d$-dim simplex is $w_0$. The appearance of a new simplex makes the weights of the $d$-dim simplex $\beta$ of the original network that share faces with the ($d$-1)-dim simplex $\alpha$ that accepted the new node change. These original weights are renewed according to the following equation

$$w_\beta \to w_\beta + b \nabla w_\beta \tag{35}$$

where $\beta \supseteq \alpha$ and $\nabla w_\beta = \dfrac{w_\beta}{s_{d,d-1}^{\alpha}}$.

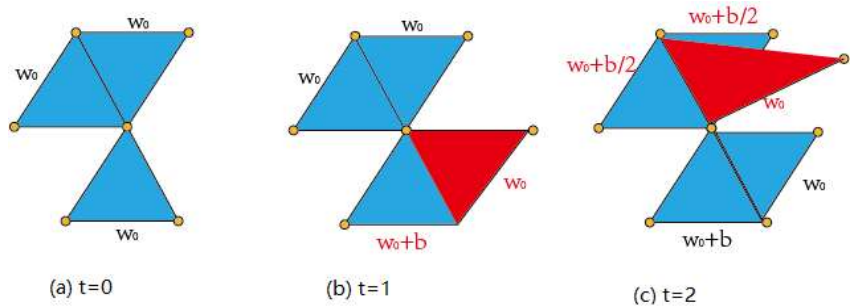

**Fig. 6.** Schematic illustration of the first two steps of the evolution for the model.

In Fig. 6, at time t=0, each 2-dimensional simplex has a weight $w_0$, and at each time step we add a new 2-dimensional simplex with a weight $w_0$. The $d$-order strength $s_{d,d-d}^{\alpha}$ of the selected $(d-1)$-dim simplex $\alpha$ changes to $s_{d,d-d}^{\alpha} + w_0 + b$.

Let $e$ be the dimension of simplex $\delta$ such that $e < d$.

When a new simplex joins the network, the $d$-order strength $s^\alpha_{d,d-d}$ of $(d-1)$-dim simplex $\alpha$ changes because of two possible scenarios: it becomes $w_0$ when $\alpha$ directly combines with a new node to form a $d$-dim simplex, or it changes due to a change in the weight of a $d$-dim simplex that has $\alpha$ as its face. Therefore, the change of the strength $s^\alpha_{d,d-d}$ satisfies

$$s^\alpha_{d,d-1}(N(t)) - s^\alpha_{d,d-1}(N(t)-1)$$
$$= m(w_0 + b)\frac{s^\alpha_{d,d-1}(N(t)-1)}{\sum_\beta s^\beta_{d,d-1}(N(t)-1)}$$
$$+ \sum_{\beta \in S_d | \beta \supseteq \alpha} m \frac{s^\alpha_{d,d-1}(N(t)-1)}{\sum_\beta s^\beta_{d,d-1}(N(t)-1)} b \frac{w_\beta}{s^\alpha_{d,d-1}(N(t)-1)}. \tag{36}$$

Factor $d$ in the previous equation reflects that each time $d$ new $(d-1)$-dim simplices are created. Hence, these new $(d-1)$-dim simplices reach the system in the form of a Poisson process. Because each time only one existing $(d-1)$-dim simplex plays its role, the second term in the previous equation does not have factor $d$. Therefore, we obtain

$$s^\alpha_{d,d-1}(N(t)) - s^\alpha_{d,d-1}(N(t)-1) = m(w_0 + 2b)\frac{s^\alpha_{d,d-1}(N(t)-1)}{\sum_\beta s^\beta_{d,d-1}(N(t)-1)}. \tag{37}$$

From the second condition of the model, it follows that

$$h^\alpha_{d,d-1}(N(t)) - h^\alpha_{d,d-1}(N(t)-1) = m\frac{s^\alpha_{d,d-1}(N(t)-1)}{\sum_\beta s^\beta_{d,d-1}(N(t)-1)}, \tag{38}$$

By substituting equation (38) into equation (37), we produce

$$s^\alpha_{d,d-1}(N(t)) - s^\alpha_{d,d-1}(N(t)-1) = (w_0 + 2b)(h^\alpha_{d,d-1}(N(t)) - h^\alpha_{d,d-1}(N(t)-1)),$$
$$s^\alpha_{d,d-1}(N(t)-1) - s^\alpha_{d,d-1}(N(t)-2) = (w_0 + 2b)(h^\alpha_{d,d-1}(N(t)-1) - h^\alpha_{d,d-1}(N(t)-2)),$$
$$s^\alpha_{d,d-1}(N(t)-2) - s^\alpha_{d,d-1}(N(t)-3) = (w_0 + 2b)(h^\alpha_{d,d-1}(N(t)-2) - h^\alpha_{d,d-1}(N(t)-3)),$$
$$\ldots,\ldots$$
$$s^\alpha_{d,d-1}(N(t_\alpha)+1) - s^\alpha_{d,d-1}(N(t_\alpha)) = (w_0 + 2b)(h^\alpha_{d,d-1}(N(t_\alpha)+1) - h^\alpha_{d,d-1}(N(t_\alpha))).$$

If you add the above equations together, you get

$$s^\alpha_{d,d-1}(N(t)) - s^\alpha_{d,d-1}(N(t_\alpha)) = (w_0 + 2b)(h^\alpha_{d,d-1}(N(t)) - h^\alpha_{d,d-1}(N(t_\alpha))), \tag{39}$$

Solving equation (27*) gives

Since $h^\alpha_{d,d-1}(N(t_\alpha)) = c_{1,m}(d-1)$, $s^\alpha_{d,d-1}(N(t_\alpha)) = w_0 c_{1,m}(d-1)$, we have

$$s_{d,d-1}^{\alpha}(N(t)) = (w_0 + 2b)h_{d,d-1}^{\alpha}(N(t)) - 2bc_{1,m}(d-1). \qquad (40)$$

By substituting equation (40) into equation (34), we obtain that when selecting ($d$-1)-dim simplex $\alpha$, existing in the network, to connect with a new node, the probability $\Pi$ of selecting depends on $\alpha$ and $d$-order degree $h_{d,d-1}^{\alpha}$ such that

$$\Pi(h_{d,d-d}^{\alpha}) = \frac{h_{d,d-d}^{\alpha} - \frac{2bc_{1,m}(d-1)}{w_0 + 2b}}{\sum_{\beta \in S_{d-1}(t)}(h_{d,d-d}^{\beta} - \frac{2bc_{1,m}(d-1)}{w_0 + 2b})}. \qquad (41)$$

Let

$$a = -\frac{2bc_{1,m}(d-1)}{w_0 + 2b}.$$

Then equation (41) becomes equation (1); and from equations (23), it follows that the cumulative distribution of $d$-order degrees of $e$-dim simplices is given by

$$P_{d,e}^{cum}(k) = \left(\frac{(a+1)(d-e)c_{1,m}(e)}{[(a+1)(d-e)-a]k + ac_{1,m}(e)}\right)^{\gamma}. \qquad (43)$$

Where

$$a = -\frac{2bc_{1,m}(d-1)}{w_0+2b}, \qquad (44)$$

$$\gamma = \frac{w_0(d+1)+2b}{\left(w_0+2b-2bc_{1,m}(d-1)\right)(d-e)+2bc_{1,m}(d-1)}. \qquad (45)$$

Substituting $a$ into equation (43) produces

$$P_{d,e}^{cum}(k) = \left(\frac{(w_0+2b(1-c_{1,m}(d-1)))(d-e)c_{1,m}(e)}{[(w_0+2b(1-c_{1,m}(d-1)))(d-e)+2bc_{1,m}(d-1)]k-2bc_{1,m}(d-1)c_{1,m}(e)}\right)^{\gamma}.$$

From equations (43), it follows that the distribution of $d$-order degrees of $e$-dim simplices is given by

$$P_{d,e}(k) = \left(\frac{(a+1)(d-e)c_{1,m}(e)}{[(a+1)(d-e)-a]k+ac_{1,m}(e)}\right)^{\gamma} - \left(\frac{(a+1)(d-e)c_{1,m}(e)}{[(a+1)(d-e)-a](k+1)+ac_{1,m}(e)}\right)^{\gamma}. \qquad (46)$$

Same as in reference [5], for $e_1 > e$, we have

$$s_{d,e}^{\beta} = \frac{1}{\binom{d-e}{e_1-e}} \sum_{\alpha \in S_{e_1} | \alpha \supseteq \beta} s_{d,e_1}^{\alpha} \qquad (47)$$

From equation (47), it follows that

$$\sum_{\alpha \in S_{d-1} | \alpha \supseteq \delta} s_{d,d-1}^{\alpha} = (d-e)s_{d,e}^{\delta} \qquad (48)$$

From equations (40), (48) and (8), we have

$$(d-e)s_{d,e}^{\delta}(N(t)) = (w_0+2b)(d-e)h_{d,e}^{\delta}(N(t)) - 2bc_{m,1}(d-1)\sum_{\alpha \in S_{d-1}|\alpha \supseteq \delta} 1. \qquad (49)$$

From equation (49) and equation (9), we have

$$s_{d,e}^{\delta}(N(t)) = \left(w_0 + 2b - \frac{2b(d-e-1)}{d-e}c_{m,1}(d-1)\right)h_{d,e}^{\delta}(N(t)) - 2b\frac{c_{1,m}(e)}{d-e}c_{m,1}(d-1). \quad (50)$$

and

$$P(s_{d,e}^{\delta}(N(t)) \leq x)$$
$$= P\left(h_{d,e}^{\delta}(N(t)) \leq \frac{d-e}{(w_0+2b)(d-e)+2b(d-e-1)c_{m,1}(d-1)}x + \frac{2bc_{1,m}(e)c_{m,1}(d-1)}{(w_0+2b)(d-e)+2b(d-e-1)c_{m,1}(d-1)}\right)$$

Denote

$$A = \frac{d-e}{(w_0+2b)(d-e)+2b(d-e-1)c_{m,1}(d-1)}$$
$$B = \frac{2bc_{1,m}(e)c_{m,1}(d-1)}{(w_0+2b)(d-e)+2b(d-e-1)c_{m,1}(d-1)}$$

So that we obtain

$$\left(s_{d,e}^{\delta}(N(t)) \leq x\right) = P\left(h_{d,e}^{\delta}(N(t)) \leq Ax + B\right). \quad (51)$$

From equation (18), it follows that

$$P(s_{d,e}^{\delta}(N(t)) \leq x) = P\left(t_\delta \geq \left(\frac{(a+1)(d-e)c_{1,m}(e)}{((a+1)(d-e)-a)(Ax+B)+ac_{1,m}(e)}\right)^{\gamma} t\right). \quad (52)$$

Without loss of generality, we may as well assume that $\delta = n(n = 0,1,2,...,)$ denote the $e$-dimensional simplex of the n*th* batch entering the network. By Poisson process theory [18], we know that the $d$-order strength distribution of the $e$-dimensional simplex is

$$P(s_{d,e} \leq x) \approx 1 - \lim_{t\to\infty} \frac{1}{EN(t)} \sum_{n=0}^{\infty} F_n\left(\left(\frac{(a+1)(d-e)c_{1,m}(e)}{((a+1)(d-e)-a)(Ax+B)+ac_{1,m}(e)}\right)^{\gamma} t\right)$$

$$= 1 - \left(\frac{(a+1)(d-e)c_{1,m}(e)}{((a+1)(d-e)-a)(Ax+B)+ac_{1,m}(e)}\right)^{\gamma}. \quad (53)$$

The $d$-order cumulative strength distribution of the $e$-dimensional simplex is given by

$$F_{2,e}^{cum}(x) = \left(\frac{(a+1)(d-e)c_{1,m}(e)}{((a+1)(d-e)-a)(Ax+B)+ac_{1,m}(e)}\right)^{\gamma}. \quad (54)$$

Hence, the density function of the $d$-order steady-state strength of $e$-dim simplices is

$$f_{d,e}^{\delta}(x) = \gamma \frac{((a+1)(d-e)-a)A}{(a+1)(d-e)c_{1,m}(e)} \left( \frac{(a+1)(d-e)c_{1,m}(e)}{((a+1)(d-e)-a)Ax+((a+1)(d-e)-a)B+ac_{1,m}(e)} \right)^{\gamma+1}, \quad (55)$$

where

$$a = -\frac{2bc_{1,m}(d-1)}{dw_0+2b}.$$

$$\gamma = \frac{w_0(d+1)+2b}{\left(w_0+2b-2bc_{1,m}(d-1)\right)(d-e)+2bc_{1,m}(d-1)}. \quad (56)$$

$$A = \frac{d-e}{(w_0+2b)(d-e)+2b(d-e-1)c_{m,1}(d-1)}. \quad (57)$$

$$B = \frac{2bc_{1,m}(e)c_{m,1}(d-1)}{(w_0+2b)(d-e)+2b(d-e-1)c_{m,1}(d-1)}. \quad (58)$$

Figs. 7-14 depict the simulated and theoretically predicted values of the weighted simplicial complex network evolution model for different initial weight $w_0$ and traffic $b$. Figs. 7 and 10 demonstrate the cumulative distribution of higher-order degrees of the weighted evolving model and the competitive evolving model with the competitiveness factor $a$ given by Eq. (44). It shows that it is reasonable to use the competitive evolving model to analyze the weighted evolving model. Figs. 11, 12, 13 and 14 show the cumulative distribution function of higher-order strengths of the weighted simplicial complex network. From Figs. 7 – 14, it can be seen that computer simulations and theoretical analyses of the weighted simplicial complex network evolution model agree with each other.

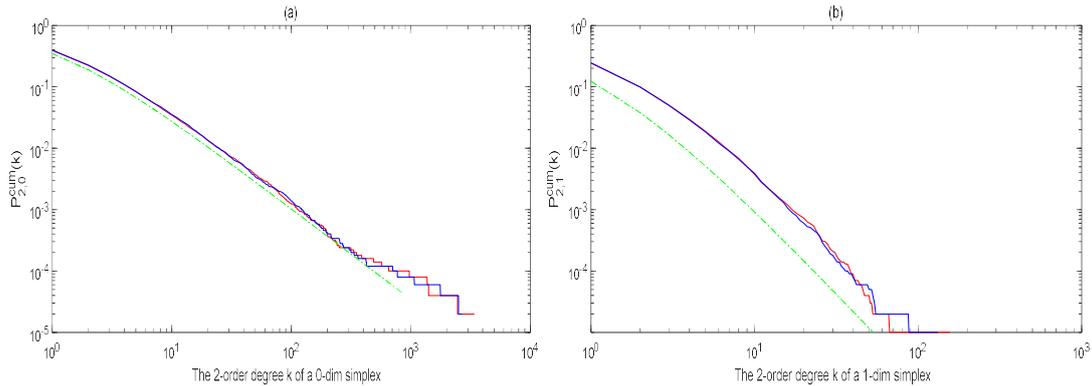

**Fig. 7. The cumulative distributions of higher-order degrees of weighted evolving model ($m=1, w_0 = 2, b = 0.1$).** The growing a network of size is N = 50,000. The lines in (a) stand for the cumulative distribution $P_{2,0}^{cum}(k)$ of the 2-order degree $k$ of the 0-dim simplex. The lines in (b) stand for the cumulative distribution $P_{2,1}^{cum}(k)$ of the 2-order degree $k$ of the 1-dim simplex. The blue lines in (a) and (b) are the simulated values of the cumulative distribution of the 2-order degree of the model. The red lines in (a) and (b) are the simulated values of the cumulative distribution of the 2-order degree

of the competitive evolving model with the competitiveness factor $a$ given by Eq. (44). The dashed lines in (a) and (b) correspond to the analytical predictions given by Eq. (43).

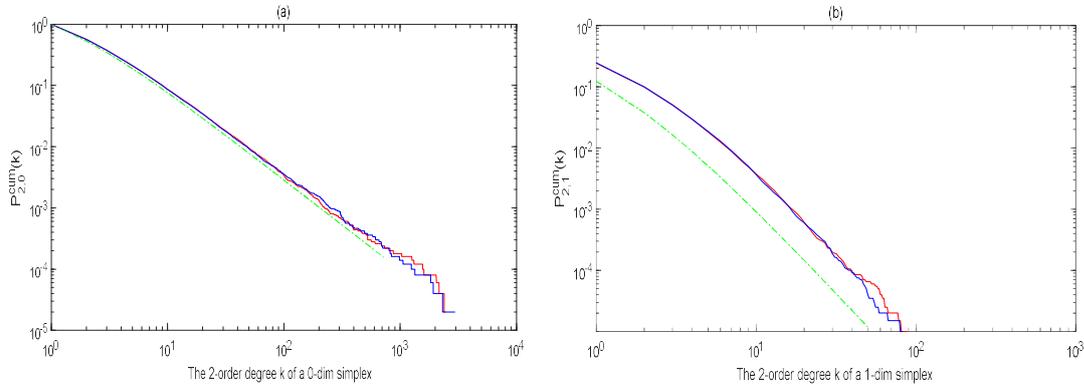

**Fig. 8**. **The cumulative distributions of higher-order degrees of weighted evolving model ($m=2, w_0 = 2, b = 0.1$).** The rest of the caption is the same as that in Fig. 7.

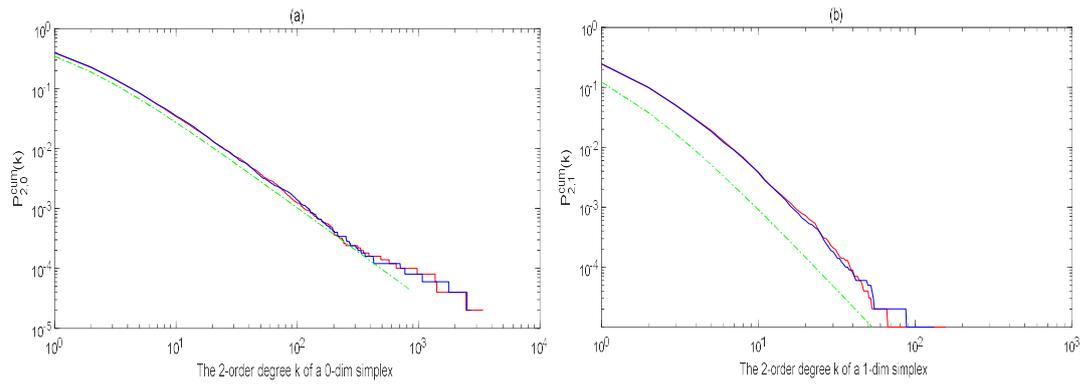

**Fig. 9**. **The cumulative distributions of higher-order degrees of weighted evolving model ($m=1, w_0 = 2, b = 0.3$).** The rest of the caption is the same as that in Fig. 7.

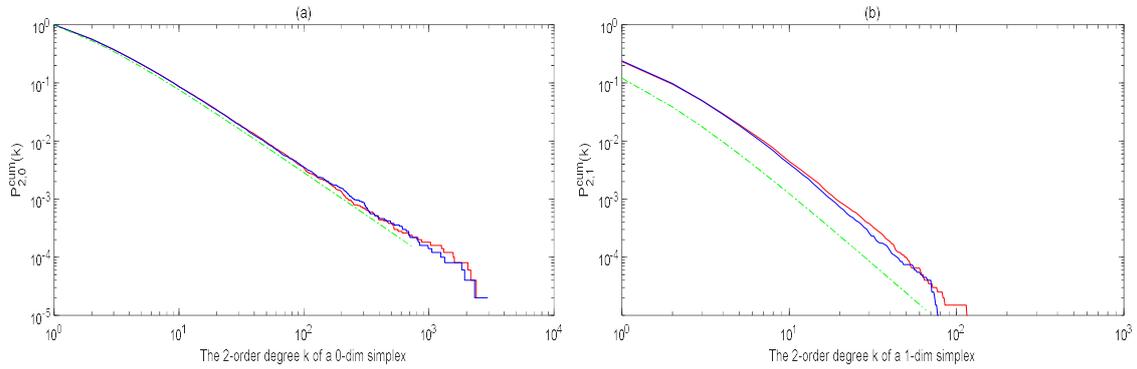

**Fig. 10**. **The cumulative distributions of higher-order degrees of weighted evolving model ($m=2, w_0 = 2, b = 0.3$).** The rest of the caption is the same as that in Fig. 7.

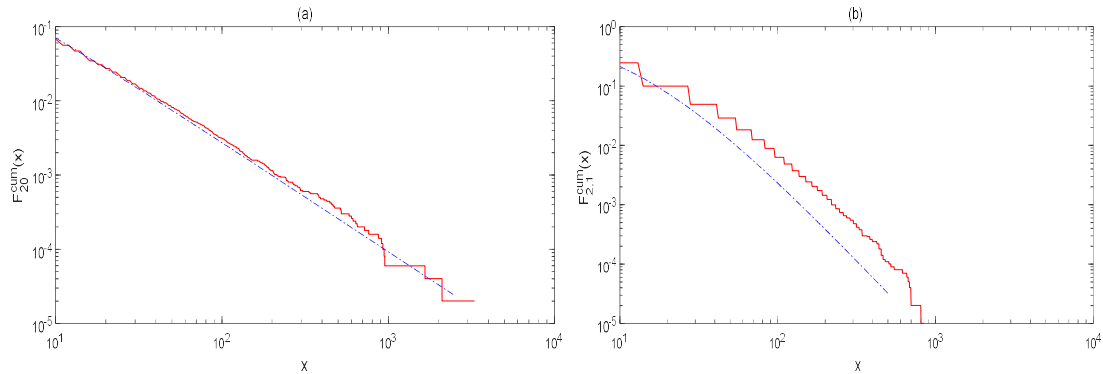

**Fig. 11. The cumulative distribution function of higher-order strength degree of Model ($m=1, w_0 = 10, b = 0.1$)**. The growing a network of size is N = 50,000. The red lines in (a) and (b) are the simulated values of the cumulative distribution function of the 2-order strength degree of the model. The dashed lines in (a) and (b) correspond to the analytical predictions given by Eq. (54).

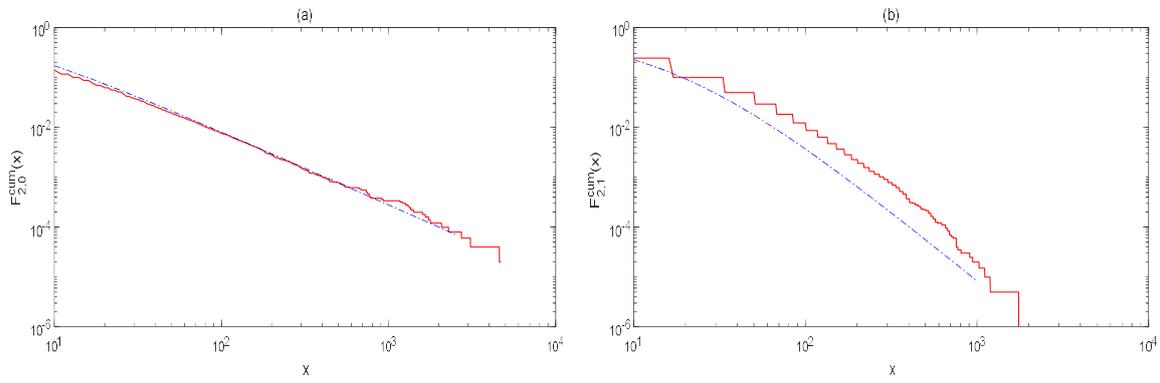

**Fig. 12. The cumulative distribution function of higher-order strength degree of Model ($m$=2, $w_0$ = 10, $b$ = 0.1)**. The rest of the caption is the same as that in Fig. 11.

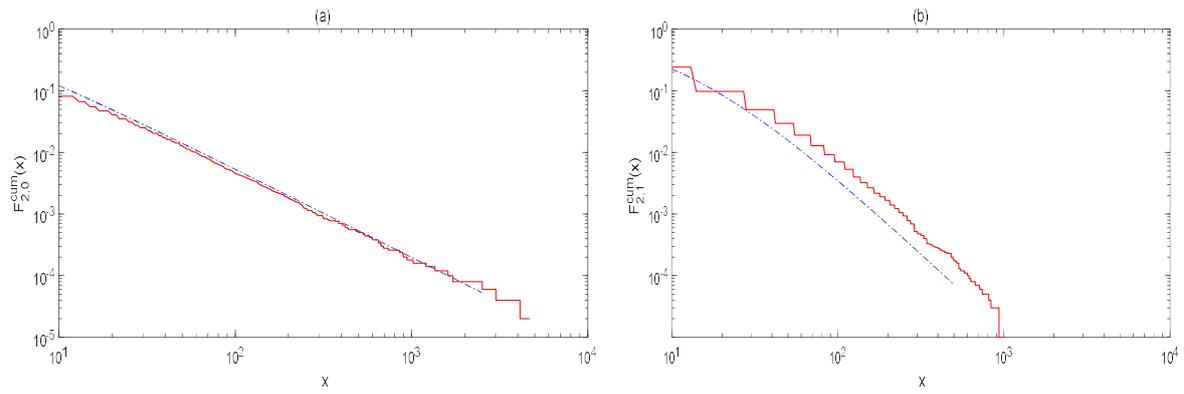

**Fig. 13. The cumulative distribution function of higher-order strength degree of Model ($m$=1, $w_0$ = 10, $b$ = 0.3)**. The rest of the caption is the same as that in Fig. 11.

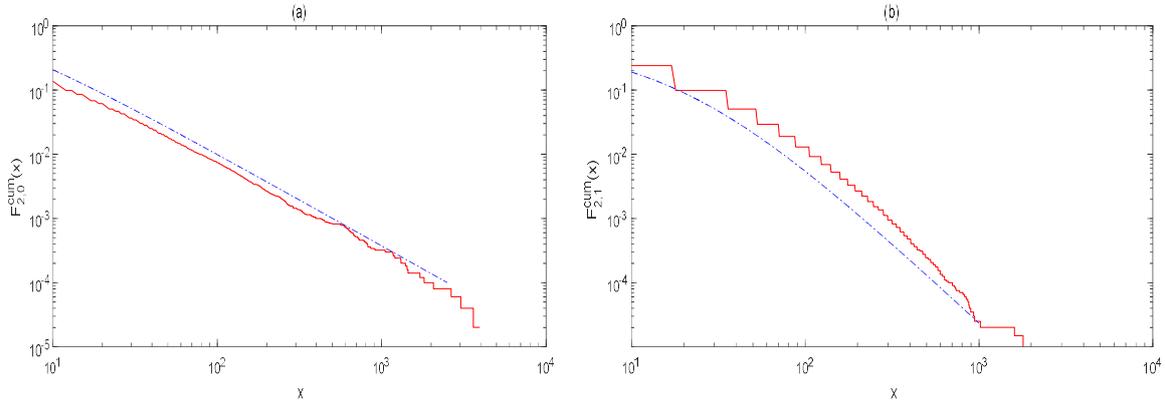

**Fig. 14.** The cumulative distribution function of higher-order strength degree of Model (*m*=2, $w_0$ = 10, *b* = 0.3). The rest of the caption is the same as that in Fig. 11.

## 5. Conclusions

In the traditional network analysis, it is assumed that the degrees of the nodes are continuous real-valued variables and the nodes are discrete to reach the system. The paper introduces the difference equation analysis approach. It avoids the assumption that the degrees of nodes are continuous in the traditional analysis network. The higher-order network competitive evolution model, introduced in this paper by employing the difference equation approach and Poisson process, is shown to be a simplicial complex model that grows in continuous time. It plays an important role in the study of higher-order networks, reflecting the emergence of the scale-free property of low-order simplexes, while the scale-free property of sub-higher-order simplices is influenced by the competitiveness factor. The results of our theoretical analysis and simulation indicate that the distribution of higher-order degrees of sub-higher-order simplices demonstrates the property of bending power-law distribution.

Considering the situation that when the simplices in a higher-order weighted network evolve, they may realistically alter the weights of neighboring simplices, we propose a weighted higher-order network evolution model. We theoretically analyze and empirically simulate this model by employing a higher-order network competitive evolution model. The results of the analysis and simulation indicate that both higher-order degree and higher-order strength of a weighted higher-order network demonstrate the scale-free property. Therefore, it can be seen that each weighted higher-order network belongs to the class of all competitive higher-order networks. Although there

arXiv:2402.06451v2are higher-order network evolution models, the research of the topological structure of higher-order networks just got off the ground. There is still a lot of works related to the evolution model of higher-order networks that are waiting to be done.

**Acknowledgments**

The authors acknowledge the supports of National Natural Science Foundation of China (Grant no. 71971139, 71571119).